\documentclass{aip-cp}

\usepackage[numbers]{natbib}
\usepackage{rotating}
\usepackage{graphicx}
\usepackage{subcaption}
\usepackage{color}
\usepackage{soul}

\begin{document}

\title{Neutrinoless Double Beta Decay \\
and High-Scale Baryogenesis}

\author[aff1]{Lukas Graf\corref{cor1}}
\author[aff1,aff2,aff3,aff4]{Julia Harz}
\eaddress{jharz@lpthe.jussieu.fr}
\author[aff1,aff5]{Wei-Chih Huang}
\eaddress{wei-chih.huang@tu-dortmund.de}

\newcommand{\AddrDortmund}{
Fakult\"at f\"ur Physik, Technische Universit\"at Dortmund, 44221 Dortmund, Germany
}

\affil[aff1]{Department of Physics and Astronomy, University College London, London WC1E 6BT, United Kingdom}
\affil[aff2]{Sorbonne Universit\' es, Institut Lagrange de Paris (ILP), 98 bis Boulevard Arago, 75014 Paris, France}
\affil[aff3]{Sorbonne Universit\' es, UPMC Univ Paris 06, UMR 7589, LPTHE, F-75005, Paris, France}
\affil[aff4]{CNRS, UMR 7589, LPTHE, F-75005, Paris, France}
\affil[aff5]{\AddrDortmund}

\corresp[cor1]{Corresponding author: lukas.graf.14@ucl.ac.uk}

\maketitle

\begin{abstract}
The constraints on baryogenesis models obtained from an observation of neutrinoless double beta decay are discussed. 
The lepton number violating processes, which can underlie neutrinoless double beta decay, would together with sphaleron processes, which are effective in a wide range of energies, wash out any primordial baryon asymmetry of the universe.
Typically, if a  mechanism of neutrinoless double beta decay other than the standard light neutrino exchange is observed, typical scenarios of high-scale baryogenesis will be excluded. This can be achieved by different methods, e.g. through the observation in multiple isotopes or the measurement of the decay distribution. In addition, we will also highlight the connection with low energy lepton flavour violation and lepton number violation at the LHC.
\end{abstract}

\section{Introduction}
The observation of a baryon asymmetry in the Universe is one of the most significant facts pointing to the need of Beyond the Standard Model (BSM) physics and any new theory of particle physics must be at least consistent with this phenomenon.
On the other hand, as we shall see later, an observation of lepton number violation at low-energy scales can shed light on the baryon number generation mechanism and also serves as a guideline for pinpointing the underlying ultraviolet complete theory. 

The observed baryon asymmetry of the Universe can be quantified as the baryon-to-photon number density ratio 
$\eta_B^{obs}=(n_{b}-n_{\overline{b}})/n_{\gamma}=\left(6.09\pm 0.06\right)\times10^{-10}$ \cite{Ade}
and various theories try to account for this value. One of the most popular scenarios of high-scale baryogenesis is so-called leptogenesis \cite{Fukugita}, which generates a $(B-L)$ number density asymmetry (where $B$ and $L$ indicates the baryon and lepton number, respectively) at some high scale, satisfying the well-known three Sakharov conditions \cite{Sakharov}: $B$ violation, $CP$ violation and departure from thermal equilibrium. The created $(B-L)$ asymmetry is then rapidly translated into the baryon asymmetry by $(B+L)$ violating sphaleron processes above the electroweak (EW) scale \cite{Kuzmin}. 

Lepton number violation (LNV) is also closely related to another observation pointing to BSM physics, namely, to the tiny, but non-zero neutrino masses and to the corresponding question of the underlying neutrino mass generation mechanism. In case that neutrinos are Dirac spinors, lepton number is conserved, which might be protected by an underlying symmetry. On the other hand, if neutrinos have Majorana masses, lepton number is violated, implying the occurrence of neutrinoless double beta ($0\nu\beta\beta$) decay \cite{Furry}.
In this report, we will highlight the fact that a non-standard mechanism which contributes to $0\nu\beta\beta$ decay will also erase a pre-existing baryon asymmetry produced at high scales in the early Universe due to the sphaleron processes. In other words, the observation of non-standard $0\nu\beta\beta$ decay will {\it falsify} high-scale baryogenesis mechanisms. The following text is mostly based on reference \cite{Deppisch4}, which can be referred to for further details.

The most relevant $0\nu\beta\beta$ decay processes are those with $\Delta L=2$ and $\Delta B=0$. Therefore, if we assume that lepton number is broken well above the EW scale, we can characterize corresponding interactions with odd-dimension $\Delta L=2$ effective operators which involve Standard Model (SM) particles only.
Up to dimension 11, there are 129 operators \cite{deGouvea}, out of which we will focus on the following four examples
\begin{eqnarray}
\mathcal{O}_5 = (L^iL^j)H^k H^l \varepsilon_{ik} \varepsilon_{jl},&& 
\mathcal{O}_7 = (L^id^c)(\bar{e^c}\bar{u^c})H^j \varepsilon_{ij}, \nonumber \\
\mathcal{O}_9 = (L^iL^j)(\bar{Q_i}\bar{u^c})(\bar{Q_j}\bar{u^c}),&& 
\mathcal{O}_{11} = (L^iL^j)(Q_k d^c)(Q_l d^c)H_m\bar{H}_i \varepsilon_{jk} \varepsilon_{lm}. \label{eq:ops}
\end{eqnarray}
Here $L=\left(\nu_L, e_L\right)^T$, $Q=\left(u_L, d_L\right)^T$, $H=(H^+, H^0 )^T$ are $SU(2)_L$ doublets and $\bar{e^c}, \bar{u^c}, \bar{d^c}$ refer to charge conjugated $SU(2)_L$ singlets. The two-component Weyl spinor notation is used: fields with~(without) a bar are right-~(left-)handed under the Lorentz group. While $\mathcal{O}_5$ is the standard Weinberg operator, the other non-standard operators will generate light neutrino Majorana masses at loop level after EW symmetry breaking.

\section{Effective Operators and Effective Couplings for $0\nu\beta\beta$ Decay}
\begin{figure*}[b!]
\centering
\begin{minipage}[t]{0.45\linewidth}
\centering
\includegraphics[scale=0.355]{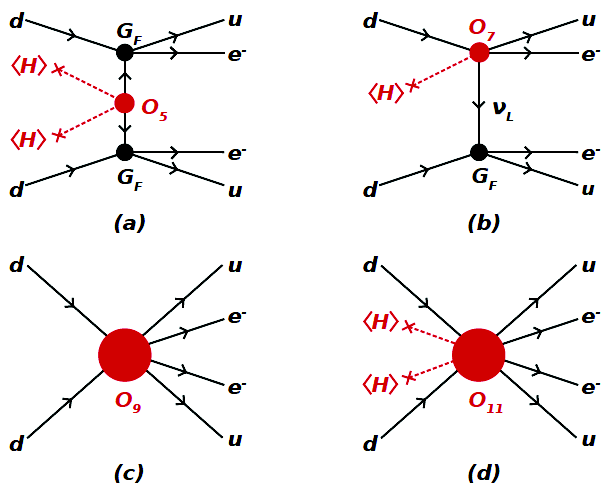}
\caption{Diagrams showing the contributions of operators (\ref{eq:ops}) to $0\nu\beta\beta$ decay in terms of effective vertices, which are point-like at the nuclear Fermi scale.}
\label{fig:1}
\end{minipage}
\quad
\begin{minipage}[t]{0.45\linewidth}
\centering
\includegraphics[scale=0.18]{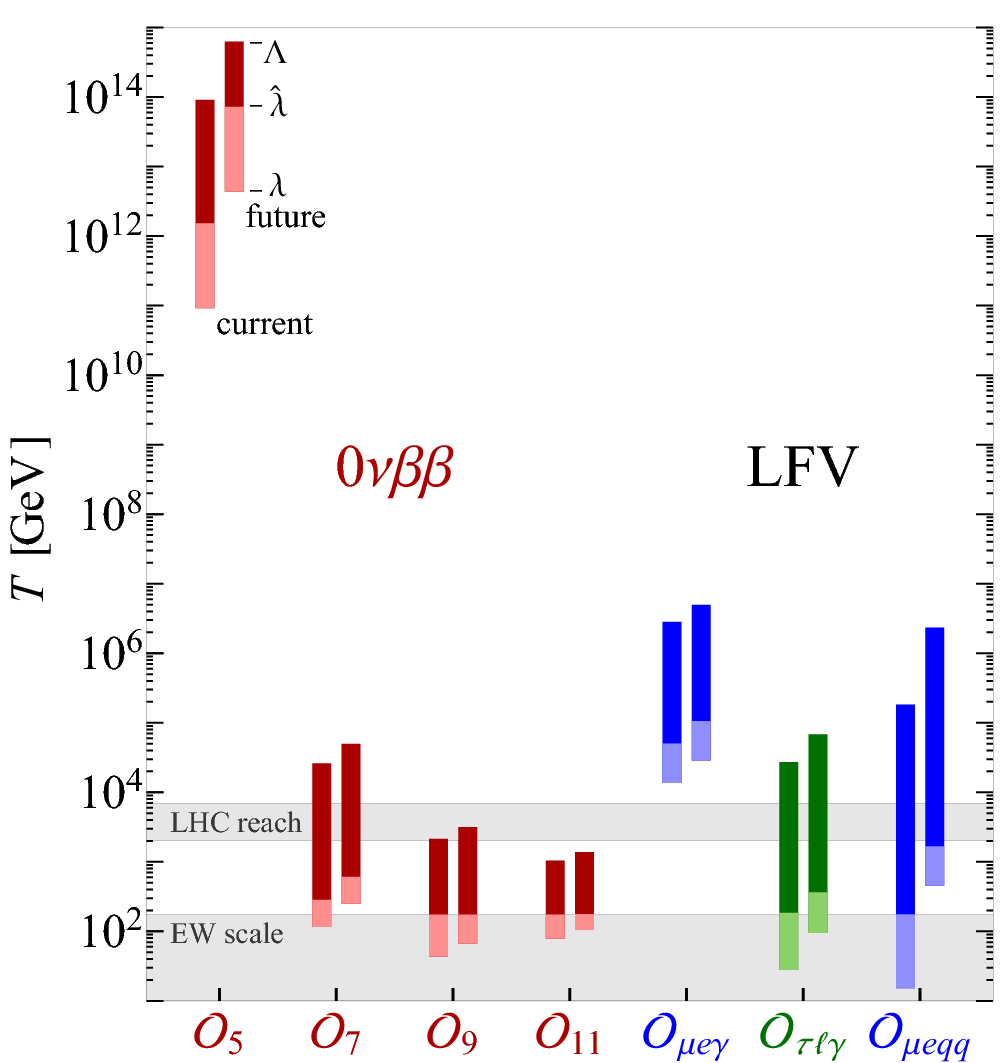}
\caption{Comparison of temperature intervals in which the LNV and LFV operators are in equilibrium assuming observation of $0\nu\beta\beta$ decay at the current/future (left/right bars) sensitivity.
}
\label{fig:2}
\end{minipage}
\end{figure*}

The non-standard contributions to $0\nu\beta\beta$ decay can be parametrized by effective operators of dimension 6 and 9~\cite{Pas}, corresponding to short-range and long-range interactions, respectively, see Fig. \ref{fig:1}. Assuming the dominance of one operator, the $0\nu\beta\beta$ decay half life is given by
\begin{equation} \label{eq:halflife}
T^{-1}_{1/2}=\epsilon_i^2G_i|M_i|^2,
\end{equation}
where $G_i$ is the nuclear $0\nu\beta\beta$ decay phase space factor and $M_i$ stands for the matrix element for a given isotope and operator. The coefficient $\epsilon_i$ is an effective coupling of a specific operator, which is related to the cut-off scale $\Lambda$ of the generating operator $\mathcal{O}_D$ in eq. (\ref{eq:ops}) as \cite{Deppisch}
\begin{equation} \label{eq:opscales}
m_e\epsilon_{5}=\frac{g^2\mathrm{v}^2}{\Lambda_5}, \ \ \ \frac{G_F\epsilon_{7}}{\sqrt{2}}=\frac{g^3\mathrm{v}}{2\Lambda_7^3},\ \ \
\frac{G_F^2\epsilon_{\{9,11\}}}{2m_p}=\Bigg\{\frac{g^4}{\Lambda_9^5},\frac{g^6\mathrm{v}^2}{\Lambda_{11}^7}\Bigg\},
\end{equation}
where v denotes the Higgs vacuum expectation value after EW symmetry breaking.
In case of the five dimensional operator, the effective coupling is simply given by the ratio of effective $0\nu\beta\beta$ decay mass $m_{ee}$ to electron mass $m_{e}$. Couplings corresponding to operators of higher dimensions are normalized by the Fermi coupling $G_F$ and the proton mass $m_p$. Moreover, the expected scaling in an ultraviolet complete model generating a particular effective operator is described by powers of a generic coupling constant $g$. However, its role is purely instructive, as in the following analysis it is set to unity for simplicity.

Using equations (\ref{eq:halflife}) and (\ref{eq:opscales}) we can obtain the scales of the effective operators $\Lambda_D$ in terms of the $0\nu\beta\beta$ decay half life and particular matrix elements, whose numerical values consequently play an important role in this calculation. The bounds on $0\nu\beta\beta$ decay half life given by experimental searches in $^{76}$Ge and $^{136}$Xe with $90\%$ C. L. currently reach values of $T_{1/2}>2.1\times10^{25}$ y \cite{Agostini} and $T_{1/2}>(1.1-1.9)\times10^{25}$ y \cite{Albert,Gando}, respectively.
Using the results on the effective couplings $\epsilon$ from \cite{Deppisch} and rescaling them with the current limits on $T_{1/2}$, one can obtain the corresponding cut-off scale $\Lambda_D$ for each operator, which are depicted in Fig. \ref{fig:2}. The planned sensitivity of future $0\nu\beta\beta$ decay searches is expected to improve by two orders of magnitude to $T_{1/2}\approx 10^{27}$ y \cite{Gomez}.

\section{Lepton Asymmetry Washout}
Let us now compute the washout effect on a pre-existing net lepton asymmetry by the above discussed effective operators. If we consider the washout driven by a single LNV  $\Delta L=2$ operator of dimension $D$, the Boltzmann equation for $\eta_L$, the net lepton number in dependence on temperature $T$ and normalized to the photon density $n_{\gamma}$, reads \cite{Deppisch4}
\begin{equation} \label{eq:boleq}
n_{\gamma}HT\frac{\mathrm{d}\eta_L}{\mathrm{d}T}=c_D\frac{T^{2D-4}}{\Lambda_D^{2D-8}}\eta_L,
\end{equation}
where $n_{\gamma}\approx2T^3/\pi^2$ is the equilibrium photon density, $H\approx1.66\sqrt{g_*}T^2/\Lambda_{Pl}$ denotes the Hubble parameter with the effective number of degrees of freedom $g_*$ ($\approx107$ in case of SM) and the Planck scale $\Lambda_{Pl}=1.2\times10^{19}$ GeV. The constant $c_D$ on the right-hand side of the equation varies for each operator and is obtaining by calculating the scattering density integrated over the whole phase space and summing over all possible initial and final states. The constants corresponding to our selection of operators (\ref{eq:ops}) are $c_{\{5,7,9,11\}}=\{8/\pi^5,27/(2\pi^7),3.2\times10^4/\pi^9,3.9\times10^5/\pi^{13}\}$.

The condition for an operator $\mathcal{O}_D$ to be in equilibrium is given by \cite{Deppisch4}
\begin{equation}
\frac{\Gamma_W}{H}\equiv\frac{c_D}{n_{\gamma}H}\frac{T^{2D-4}}{\Lambda_D^{2D-8}}\approx 0.3c_D\frac{\Lambda_{Pl}}{\Lambda_D}\left(\frac{T}{\Lambda_D}\right)^{2D-9} \gtrsim 1.
\end{equation}
This relation is satisfied whenever the temperature $T$ lies in the interval
\begin{equation}
\Lambda_D \gtrsim T \gtrsim \lambda_D\equiv\Lambda_D\left(\frac{\Lambda_D}{c'_D\Lambda_{Pl}}\right)^{\frac{1}{2D-9}}.
\label{ineq1}
\end{equation}
The lower limit $\lambda_D$ therefore represents the temperature above which any pre-existing lepton number asymmetry will be washed out in case that $0\nu\beta\beta$ decay is observed at the corresponding rate and if the given operator $\mathcal{O}_D$ is the dominant contribution. On the other hand, the scale $\Lambda_D$ indicates the upper limit given by validity of the effective operator approach. Above this scale the underlying (UV-completed) model must be considered.

If we consider a particular operator, e.g. $\mathcal{O}_7$, we can rewrite the inequalities (\ref{ineq1}) using (\ref{eq:opscales}) and (\ref{eq:halflife}) in terms of $0\nu\beta\beta$ decay half-life; i.e.,
\begin{equation}
\left(\frac{T_{1/2}}{10^{25}\mathrm{y}}\right)^{\frac{1}{6}} 2.3\times 10^{4}\ \mathrm{GeV} \gtrsim T \gtrsim \left(\frac{T_{1/2}}{10^{25}\mathrm{y}}\right)^{\frac{1}{5}} 98.9\ \mathrm{GeV},
\end{equation}
where $T_{1/2}$ can be substituted by a chosen value of $0\nu\beta\beta$ decay half-life or by its future limit. If we set $T_{1/2}$ to the (approximate) current experimental limit $10^{25}$ y, then these inequalities yield simply the expression (\ref{ineq1}) for $D=7$ with corresponding numerical values of the involved scales.

A more precise determination of the lower limit for the washout can be obtained by solving the Boltzmann equation (\ref{eq:boleq}). In this way we can determine the scale $\hat{\lambda}_D$, above which the lepton asymmetry washout induced by a particular operator is effective enough to suppress the primordial asymmetry down to the EW scale, where the remaining difference is translated into a baryon asymmetry by sphaleron processes. This more stringent limit reads
\begin{equation}
\hat{\lambda}_D\approx\left[\left(2D-9\right)\log_e\left(\frac{10^{-2}}{\eta_B^{\mathrm{obs}}}\right)\lambda_D^{2D-9}+\mathrm{v}^{2D-9}\right]^{\frac{1}{2D-9}}.
\end{equation}
The considered primordial asymmetry is of order one. The intervals of effective washout for all the operators are depicted in Fig. \ref{fig:2} for both the current and future $0\nu\beta\beta$ decay sensitivities. 
The most obvious feature of the obtained results is the big gap between the Weinberg operator ($\approx 10^{14}$ GeV) and other LNV operators ($\approx 10^{3-4}$ GeV), which means that $0\nu\beta\beta$ decay can provide information about both low and high scales. On the one hand, if $0\nu\beta\beta$ decay is observed and triggered by any of the non-standard mechanisms, any high-scale baryogenesis scenario above corresponding $\hat{\lambda}_D$ is excluded. Nevertheless, if $0\nu\beta\beta$ decay is dominated by the standard mass mechanism, the origin of neutrino masses and baryogenesis are most probably high-scale phenomena. 

Hence, it turns out that it is very important to be able to distinguish non-standard mechanisms from the standard one since they correspond to very different cut-off scales, leading to very different temperature ranges within which the wash-out effect on lepton asymmetry is efficient. Naturally, the observation of $0\nu\beta\beta$ decay alone would not be enough to pin down the underlying effective operator. However, different strategies exist to distinguish different mechanisms of $0\nu\beta\beta$ decay (for further discussion we refer to \cite{Deppisch4}). Possibilities are, for example, to observe the decay from different isotopes \cite{Deppisch2} or to measure the angular and energy distribution of the outgoing electrons, which could be achieved by the SuperNEMO experiment \cite{Arnold}. This is applicable for the operator $\mathcal{O}_7$, whose final state contains electrons of opposite helicities. On the other hand, if the effective operator $\mathcal{O}_9$ or $\mathcal{O}_{11}$ contributes most, they could be probed at the LHC. In fact, observation of LNV at the LHC would already imply an exponential reduction of any primordial lepton asymmetry and consequently an exclusion of highscale leptogenesis models~\cite{Deppisch3,Deppisch5}.


Unlike resonant processes at LHC, $0\nu\beta\beta$ decay can probe LNV only in the first lepton generation. However, if LFV is observed, our argumentation can be extended to the lepton asymmetry washout on other flavours. This is shown in Fig.~\ref{fig:2}. For the current and estimated future constraints from $\mu\rightarrow e\gamma$, $\tau \rightarrow \ell \gamma$ and $\mu - e$ conversion in nuclei, the corresponding temperature interval is shown where all flavours are in equilibrium. Thus, an overlap among the intervals of LFV and LNV processes ensures that the net numbers of both involved flavours are efficiently washed out.

To conclude, we have described a tight connection between low-energy LNV and the observed baryon number asymmetry. The observation of $0\nu\beta\beta$ decay triggered by a non-standard mechanism together with an observation of LFV would exclude high-scale baryogenesis scenarios. However, some caveats should be taken into account. For instance, the baryon asymmetry can be generated below the electroweak scale or certain mechanism protecting the asymmetries from washouts can exist.
In any case, we wish to encourage experiments to perform extensive searches of any LNV processes, as their discoveries can lead to fundamental conclusions about the early stages of the Universe formation.


\section{Acknoledgements}
We would like to thank co-authors of the original paper \cite{Deppisch4}, Frank F. Deppisch, Martin Hirsch and Heinrich P\"{a}s for corrections and useful discussions. The work of JH and WCH was supported partly by the London Centre for Terauniverse Studies, using funding from the European Research Council via the Advanced Investigator Grant 267352.


\nocite{*}
\bibliographystyle{aipnum-cp}%

\end{document}